\documentclass[12pt,english]{article}

\usepackage{graphicx}
\usepackage{amsmath}
\usepackage{natbib}
\bibpunct{(}{)}{,}{a}{}{;}
\usepackage[top=1in, bottom=1in, left=1in, right=1in]{geometry}
\usepackage{setspace}

\title{Large-scale study of social network structure and team performance in a multiplayer online game
}

\author{Antti Ukkonen\\Department of Computer Science\\University of Helsinki, Helsinki, Finland\\\texttt{antti.ukkonen@helsinki.fi}
  \and
  Juho Hamari\\University of Tampere, Tampere, Finland\\\texttt{juho.hamari@uta.fi}}

\newcommand{\degree}{d}
\newcommand{\mindeg}{\degree_{\min}}
\newcommand{\maxdeg}{\degree_{\max}}
\newcommand{\dens}{e}

\newcommand{\win}{w}
\newcommand{\radi}{\textsf{Radiant}}
\newcommand{\dire}{\textsf{Dire}}
\newcommand{\both}{\textsf{Both}}

\renewcommand{\cite}{\citep}

\newif\iftable
\newif\iffigure
\tabletrue
\figuretrue

\begin{document}

\maketitle
\begin{abstract}
A question of interest in both theory and practice is
if and how familiarity between members of a team,
expressed in terms of social network structure,
relates to the success of the team in a given task.
In this paper we revisit this important question
in a novel manner
by employing game outcome statistics from Dota 2,
a popular team-based multiplayer online game,
combined with network data from Steam Community,
a social networking service for gamers.
We conduct a large-scale analysis of 4168 teams
to study how network density,
and the minimum and maximum degree of
the within-team social network
are associated with team performance,
and determine how this association is moderated by team skill.
We observe that
minimum degree is strongly associated with good performance,
especially in teams with lower skill.
Together with previous results on network density that we corroborate
in this paper,
our findings suggest that
a successful team is not only moderately connected overall,
but its members
should also individually
have not too few nor too many within team connections.

\end{abstract}


\section{Introduction}
Teamwork is prevalent in various occupations, such as health care (e.g. surgery teams), transportation (e.g.~airline cabin crews), sports/eSports, school work (e.g.~ad-hoc study groups), and knowledge work (e.g.~R\&D teams). In all these fields, a crucial topic of interest is team performance, and how it can be improved. Team performance in general is affected by a number of factors, such as the organisational environment in which the team operates \cite{thamhain2004linkages}, individual team members’ cognitive abilities \cite{devine2001smarter} and dispositions and skills \cite{stewart2006meta}, as well as interpersonal relations and trust between team members \cite{de2003task}. In this paper we focus on the two latter factors, by revisiting the question of {\em how social network structure within a team affects its performance}, and how this is {\em moderated by the overall skill level} of the team.

Intuition suggests that when team members are more familiar with each other, they also communicate and collaborate more efficiently, and consequently achieve higher performance. Therefore, a number of previous studies have considered the effects of a teams {\em density}, i.e., the fraction of dyadic ties it contains, on performance.
These studies have found a positive effect of network density (or simple variants thereof) on performance in a variety of contexts, such as R\&D \cite{baldwin1997social,reagans2001networks,huckman2009team}, healthcare \cite{reagans2005individual,elbardissi2008identifying}, education \cite{de2014strength}, as well as sports \cite{grund2012network}. A meta-analysis by \citet{balkundi2006} of 37 studies involving 3098 teams in total supports these findings.
But on the other hand, a number of studies have also observed that very high levels of within team connections may also may have detrimental effects on performance \cite{katz1982investigating,berman2002tacit,kratzer2004stimulating,OhEtAl2004,uzzi2005collaboration,guimera2005team}. This phenomenon is usually attributed to the teams being static and developing stagnant working routines that result in inferior performance, or failures to innovate as novel ideas are often introduced by new team members.

Previous literature has thus found support both for
a positive effect between density and performance, as well as
a nonlinear relationship where further increases in density
tend to weaken performance.
Of course studies that only test for linear relationships cannot find nonlinear effects even if they exit.
Moreover, especially with small networks,
density may not be an ideal measure to study within team social dynamics,
as we discuss below in Section~\ref{sect:metrics} in more detail.

\begin{figure}[t]
\centering
\iffigure
\includegraphics[width=\columnwidth]{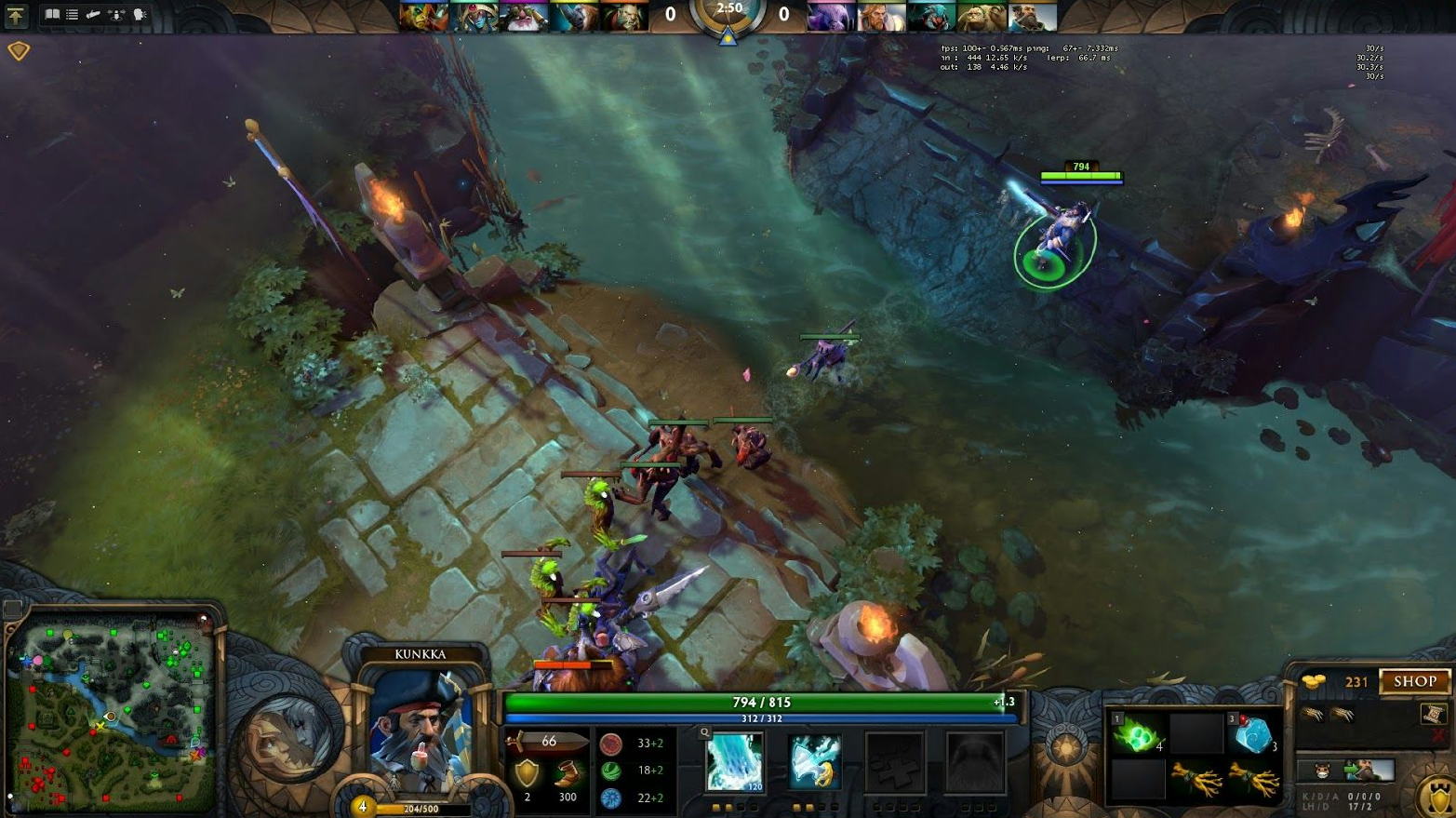}
\fi
\caption{Screenshot of Dota 2. The map as well as other user interface components are shown in the bottom part of the screen. Each player controls one of the avatars.}
\label{fig:dota2gameplay}
\end{figure}

Hence, we take an approach that
combines novel, large-scale data with
an exploratory yet statistically sound analysis technique that
also considers the {\em minimum and maximum degree} of a network
in addition to its density.
In concrete,
we collected performance data for 4168 teams
from Dota 2 \cite[see also Figure~\ref{fig:dota2gameplay}]{dota2game},
one of the most popular online games at the moment,
and joined this with network data from Steam,
a social networking service for gamers.
This methodology allows to control for a number of issues that
may affect team performance studies in general.
Namely, in the real-world
1) teams vary in size as well as skill,
2) the tasks are often not precisely identical,
3) the contexts may differ and be difficult to control,
4) the tasks or projects vary in length,
5) teams might be formed by different mechanics, and finally,
6) the number of teams available for a study (e.g. from a single organisation) can be fairly low.

We argue that
by using data from a multiplayer online game
we can very efficiently overcome these limitations
to conduct an investigation of team performance in
a realistic, large-scale, yet controlled environment,
where the teams are carrying out exactly the same task.
At the time of writing, Dota 2 is the largest eSports game in the world \cite{hamari2017esports} with thousands of matches being played daily. In this game two teams of five members each play against each other in matches of (approximately) 30 minutes in length. Moreover, much like sports teams, the teams in Dota 2 need both skilled individuals, as well as good group dynamics to be successful.
This makes Dota 2 an ideal platform to study team performance
in a context that nicely addresses all six issues mentioned above.
In particular, in Dota 2 all teams
1) consist of exactly five members having roughly equal skill,
2) have one and the same goal,
3) organize and act in exactly the same environment,
4) the length of the activity is the same for all teams,
5) all teams are formed in the same manner, and finally
6) collecting large quantities of data is easy.

In this paper we focus on
the social network between team members, called ``team network'' below.
We assume that an edge between two members in the team network signals the presence of pre-existing, voluntary, informal, self-reported interactions between the team members. The network does thus not reflect external structures, such as fixed communication channels, or supervisor-subordinate relationships. For every member of the team, the other members are assumed to be either prior acquaintances or strangers. Our definition of a dyadic tie can be thus understood as being {\em expressive rather than instrumental} \cite{lincolnmiller79}.

\subsection{Density, minimum and maximum degree, skill level}
\label{sect:metrics}
Network science literature is abundant with various statistics to characterise network structure, such as the average shortest-path length or the clustering coefficient \cite{wasserman1994social}.
Also other complex characterisations of the team network such as core-periphery structures \cite{borgatti2000models} and structural holes \cite{burt2004structural} have been considered, and are certainly meaningful for team performance, see for instance \cite{cummings2003structural}.
However, the above measures are mainly intended to address global properties of a possibly very large network, and are hence less suitable for very small networks, such as the ones studied here.
Hence, we have chosen to use two simple and intuitive metrics of the team network:
{\em density}, {\em minimum degree} and {\em maximum degree}.
Not only are these easy to explain and understand,
but they are also simple enough for practical managment settings
when assembling teams.

Density is a standard measure used in almost all previous studies on
network structure and team performance,
and it captures the overall connectedness of a team.
Since all of our networks contain precisely five nodes,
we can express density simply as
the number of edges in the network,
with ten being the maximum.

On the other hand, node degree, i.e., {\em the number of connections a team member has in the team network}, is another simple way to quantify network structure. In the context of teams, node degree has been used previously to address questions e.g.~about success of enterprise system adoption \cite{sasidharan2012effects}, effects of skill on team performance \cite{devine2001smarter}, software engineering practices \cite{zanetti2013categorizing}, as well as tie strength \cite{de2014strength}.
The minimum and maximum degrees are simply the degrees of the least and most connected team members, respectively.

\begin{figure}
\centering
\includegraphics[width=0.5\columnwidth]{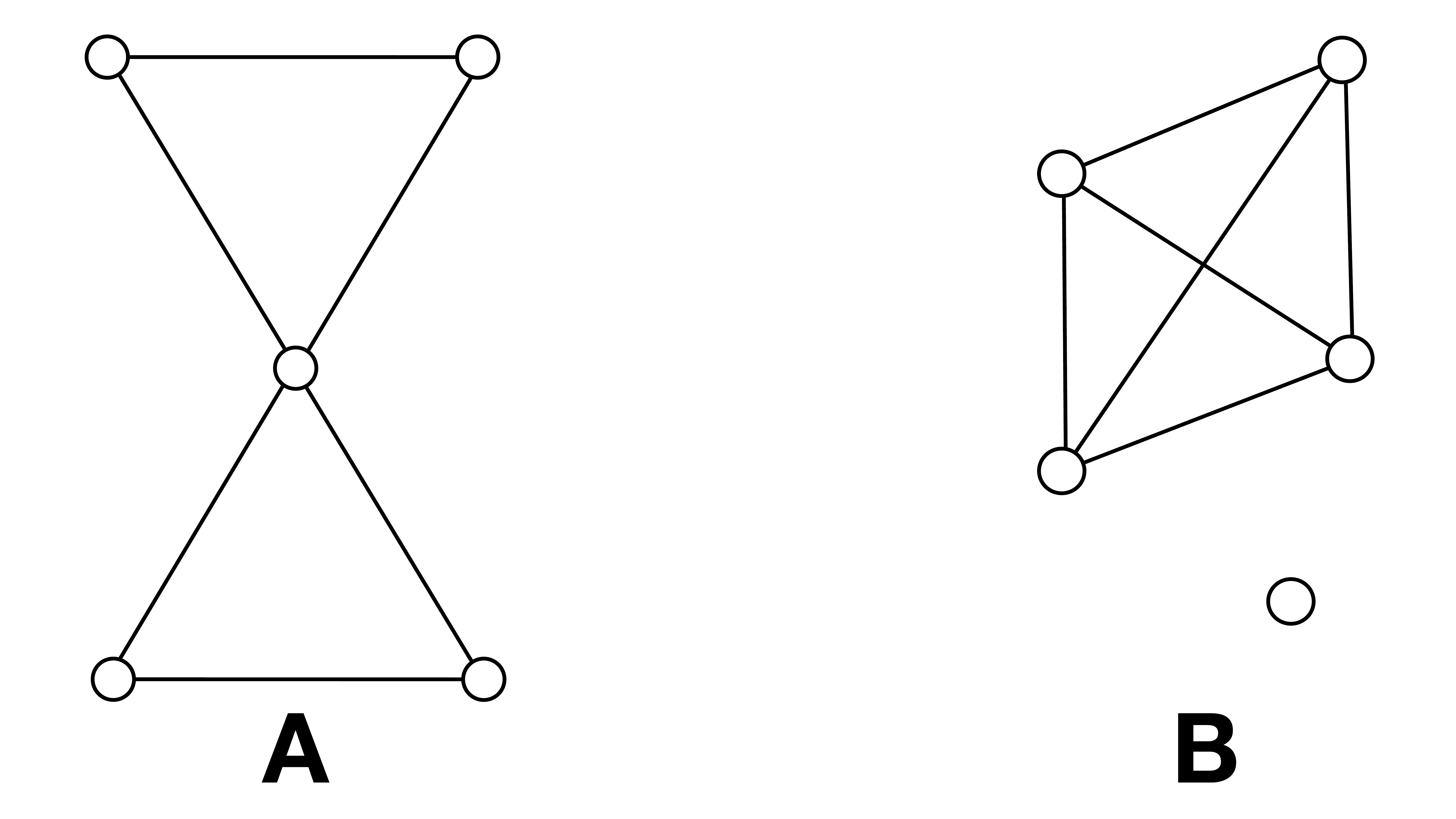}
\caption{Two social networks, A and B, both having five nodes and six edges and thus equal density, but very different structure. Note that network A has minimum and maximum degrees of 2 and 4, while for B these numbers are 0 and 3.}
\label{fig:intro_example}
\end{figure}
Minimum and maximum degree are measures complementary to density.
To see this, consider the two team networks shown in Figure~\ref{fig:intro_example}, both having exactly five members, of which team A consists of {\em two triangles that share one node}, while team B is structurally a {\em clique of four nodes together with a single isolated node}.
Both networks have exactly six edges, and hence the same density, but the networks are structurally extremely different. It does not seem at all obvious that these structures would affect team performance in the same manner.
On the other hand, the minimum and maximum degrees of the first network are 2 and 4, respectively,
while in the second network these numbers are 0 and 3.
The minimum and maximum degree thus capture structural properties
that are not taken into account when considering only density\footnote{Note that {\em average degree} is in fact equivalent to density up to a constant factor.}.

Furthermore, skill and experience in the activity the team is engaged in have as such a positive effect on team performance \cite{neuman1999team,humphrey2009developing}.
And, as the team members become more skilled,
performance may become less affected by network structure \cite{balkundi2006}.
This has been considered in previous literature to a lesser extent,
possibly partly due to the difficulty in collecting suitable data.

A further advantage of using an online game to study team performance is that controlling for the overall skill level of a team is very easy.
To provide an enjoyable gaming experience, the Dota 2 platform aims to always pair equally skilled teams against each other. Every team in our data is associated to one of three skill levels that is determined automatically by the game platform based on previous gameplay results of the teams members. (The precise details of this mechanism are proprietary to the game designers.) The outcome of a single match is thus not affected by substantial differences in overall skill level of the two teams. By comparing matches of low skilled teams to matches of teams having high skill, {\em we can study to what extent skill moderates effects that network structure has on performance}.

\subsection{Research question and basic methodology}
In summary,
we employ (to the best of our knowledge)
the largest (N=4168) real-world data
about network structure and team performance to date
\begin{itemize}
\item[a)] to replicate results of existing studies about effects of network structure on performance in a novel setting,
\item[b)] to study the association between minimum/maximum (within team) degree and performance, and
\item[c)] to determine if team skill level moderates the above effects.
\end{itemize}
We argue that our study has high internal validity,
because of the systematic nature of the Dota 2 game.
In particular, all teams are of equal size (5 members),
and have the same time ($\sim$30 minutes) to
accomplish the same objective (reach the game objectives).

Team performance is operationalised by
a binary variable indicating if the team won or lost a match,
while network structure is captured by
the three statistics described above:
density, as well as minimum and maximum degree.
This study is exploratory in nature.
Our basic methodology is to
estimate {\em expected winning probabilities}
for different types of team networks
(as characterised by the network statistics)
and skill levels, 
and compare these
against probability estimates from
a statistical {\em baseline model}
where {\em associations between network structure and performance
have been explicitly removed}.
This approach allows us to identify types of networks
in which structure indeed is associated with
a significant decrease or increase in performance.
Finally,
there are some subtle issues with data collection and preparation
that our analysis technique must carefully take into account.
These are discussed in more detail below.

\section{Materials and methods}
One of the contributions of this paper is to
demonstrate the use of combining real-world game data with
a separate social network for team performance studies.
This presents some methodological challenges related to
data collection, preparation, and analysis.
In short,
we retrieved publicly available Dota 2 match statistics from a website about game outcomes (dotabuff.com), and joined these with publicly available contact lists of the players from the Steam social network (steamcommunity.com).
Both the match statistics as well as the contact lists
are obtained by downloading publicly accessible web pages from the Internet,
and parsing the relevant information from these with an automated script.
We first give an overview of Dota 2 and Steam,
then describe data collection and preparation,
and conclude this section by discussing
technical issues that our data presents for analysis,
and how we tackle these with a simple but statistically robust analysis technique.

\subsection{Dota 2}
Dota 2 \cite{dota2game} is a so called MOBA (Multiplayer Online Battle Arena) game produced by Valve Corporation with over 10 million active players at the time of writing.
In MOBAs players play in discrete {\em matches} involving two teams of a few players.
In Dota 2 the two teams are called ``Radiant'' and ``Dire'',
and there are {\em exactly five} players in both teams.
Dota 2 and other MOBA games are commonly played from an isometric
perspective on a symmetric map, where every player controls a single
character as shown in Figure~\ref{fig:dota2gameplay}. Both teams
occupy a stronghold at opposite corners of the map. The objective is
to destroy the opposing team's main structure(s) as well as other
buildings (marked as green and red dots on the map seen in the lower
left corner of Figure~\ref{fig:dota2gameplay}).
The first team to reach this objective is the winner.

Before a match begins,
the teams are formed in an ad-hoc manner by a mechanism that
aims to assign roughly equally skilled players together.
Moreover, the matches are balanced in the sense that
both teams are approximately of the same skill level
There are three skill levels, called ``normal'', ``high'', and ``very high''.
See also \ref{app:dota2matchmaking}.
In the matches that we consider the teams are transient.
That is, a certain set of five players
play together in very few matches, and
most teams participate in only a single match.

\subsection{Steam and the Steam Community social network}
Steam (www.steam.com) is an online service for gamers provided by Valve Corporation. It is both a retail marketplace for game developers, as well as a social networking service (similar to e.g.~Facebook or Twitter) for players. Steam is the largest digital distribution platform for PC games with 1,800 titles for sale and 35 million active users \cite{steamstats}. In practice almost everyone who actively plays games on a PC is bound to use Steam at least to purchase games.

To use Steam, a player must create a digital identity called a {\em Steam profile}. Similar to other social networking sites, the players can connect with each other by adding the profiles of other Steam users to their list of friends.
This social network can be accessed at steamcommunity.com. 
The network is symmetric, and a connection is formed only when both players choose to accept this. Importantly, players are never automatically connected on the Steam platform.
The semantics of a connection in the network varies. Players may connect because they know each other in the real-world, but many connections are between players who have only met online.

\subsection{Data collection}
Steam has a built-in data collection mechanism that is based on an opt-in system for collecting game statistics from all players who have agreed to share their data. In case of Dota 2, the statistics provide information about the outcome, player identities, and various other game related parameters for every match. For matches where at least one of the players had decided to upload his information, basic statistics about the match appear at various websites. One such website is dotabuff.com. We wrote a simple computer program that periodically polls the dotabuff.com website, and downloads detailed statistics of every public Dota 2 match that appears on the site. This data was joined with public contact lists of the Steam social network for those players for which profile information was available.
Data was collected from November 13, 2014 until January 5, 2015.
(Also see \ref{app:scraping}.)
This raw data contains statistics for 93158 public Dota 2 matches.
While this sounds like a large number,
most of the matches must be discarded due to part of the data missing,
as we discuss next.

\subsection{Data preparation}
For further analysis we only keep matches that used ``normal matchmaking'' to build the teams, and are based on the ``all pick'' game mode (see Appendix Section~\ref{app:dota2modes}). These are the most common types of matches, and our data contains 76174 of them (81.8\% of the raw data). Matches of this type commonly last for approximately 30 minutes.

We then constructed the {\em team networks} for both teams.
If two team members appear on each others contact lists in Steam, we inserted the corresponding edge into the team network. (Note that the two networks are always disjoint. We only consider within-team connections, even if between-team links might exist.)
This can only be done when
both team members have opted-in to the data collection process,
otherwise their Steam identifiers are not available in the match statistics obtained from dotabuff.com.
As it is impossible to match anonymous players to their Steam profiles at steamcommunity.com,
the team network remains incomplete if some players have not revealed their profiles.

To construct the complete team network, we can therefore only consider matches where the Steam profile identifier of all five players in at least one of the teams is known.
See Table~\ref{table:known_counts} for a breakdown of the matches in terms of the number of non-anonymous players for both Radiant and Dire.
Only matches from the bottom row and the rightmost column in Table~\ref{table:known_counts} can thus be used (shown in bold, 3822 matches in total, 4.1\% of the raw data).

From these matches we derive three datasets,
\radi, \dire\ and \both, for the remaining analysis.
The dataset \radi\ (2186 teams) consists of those team networks where where all
five Radiant team members are non-anonymous.
The dataset \dire\ (1972 teams) is constructed in the same manner using Dire teams. The third dataset \both\ (4168 teams) is simply the combination of \radi\ and \dire.
Most of the analysis concerns the dataset \both,
but by considering the Radiant and Dire teams also separately,
we can conveniently compare two subpopulations
where possible associations between network structure and performance
should remain the same.
Indeed, there is no prior reason to assume
any differences between \radi\ and \dire\
in this sense.
For every team in every dataset we of course also know
if the team won the match in which it played.

\iftable
\begin{table}
\centering
\caption{Breakdown of the number of observed matches in terms of the numbers of known player identities in both teams.}
\label{table:known_counts}
{\small
\begin{tabular}{rrrrrrr}
 & \multicolumn{6}{c}{Dire} \\
 Radiant  &    0  &  1  &  2  &  3 &   4 &   {\bf 5}\cr
\hline
  0 & 20557 & 7218 &  2879 &  1183 &   444 & {\bf 142} \cr
  1 & 7286  & 4951 &  2847 &  1511 &   718 & {\bf 232} \cr
  2 & 2980  & 3030 &  2380 &  1702 &   931 & {\bf 308} \cr
  3 & 1249  & 1641 &  1773 &  1601 &  1075 & {\bf 439} \cr
  4 &  504  &  762 &  1059 &  1124 &   947 & {\bf 505} \cr
 {\bf 5} &  {\bf 117}  &  {\bf 253} &  {\bf 372} & {\bf 514} & {\bf 594} & {\bf 346} \cr
\hline
\end{tabular}
}

\end{table}
\fi

\subsection{Network statistics and winning probability}
We continue with basic definitions
of our operationalisations of network structure
and team performance.
The {\em density} $\dens$ of a team network
is the number of edges it contains.
We group the
edge density $\dens$ to six buckets
that consist of
$0$, $1-2$, $3-4$, \ldots, $9-10$ edges.
The {\em degree} of a player is
the number of neighbours the player has in the team network.
With teams of five members,
the degree ranges from zero (0) to four (4).
The minimum and maximum degrees of a team,
denoted $\mindeg$ and $\maxdeg$,
are the degrees of its least and most connected members.
In practice each dataset can be viewed as having four variables
that are $\dens$, $\mindeg$, $\maxdeg$, and
a binary variable $\win$
that indicates if the team won the match.

Our objective is to study how different values of
$\dens$, $\mindeg$, and $\maxdeg$
are associated with team performance.
In the subsequent analysis
we thus consider the {\em winning probability} $\Pr(\win)$ of a team,
{\em conditioned} on some value of either $\dens$, $\mindeg$, or $\maxdeg$.
We write
$\Pr(\win\mid x = i)$,
where $x$ is either $\dens$, $\mindeg$, or $\maxdeg$
to denote the winning probability in the condition $x=i$.
For example,
$\Pr(\win\mid\mindeg=3)$
denotes the winning probability of a team
in the condition
where the least connected player has
exactly three within-team connections.
Every condition corresponds thus to some specific type of network structure.

To compute the estimate of
$\Pr(\win\mid\mindeg=3)$
in a given dataset,
we consider all teams
with $\mindeg=3$ and
compute the fraction of teams that won.
Likewise for other conditions.
These estimates are compared
against the {\em baseline probability of winning}, denoted $\Pr(\win)$
and defined simply as the fraction of winning teams in the entire data.
In short,
our estimate of $\Pr(\win \mid x = i)$ should be
``substantially smaller or larger''
than $\Pr(\win)$ for the estimate to be
indicative of an association between
networks belonging to condition $x=i$ and performance.
Details of this procedure are described next.

\subsection{Resampling procedure}
In the absence of prior assumptions on team performance and confounding factors,
a team should have a fifty-fifty chance of winning.
Meaning, without any condition on network structure we should observe
a baseline probability of $\Pr(\win) = 0.5$ for a team to win,
because in every match there are two teams,
and one of these always wins (there are no ties).
However, our baseline estimate of the winning probability
is in practice distorted by two factors.

First, a known property of Dota 2 is that
in matches of approximately 30 minutes in length,
the team playing Radiant has a slightly higher chance of winning\footnote{In
the player community this is usually attributed to some subtle game design details (e.g.~orientation of the game map) that may provide a tiny advantage to Radiant.}.
As there are more Radiant teams in \both,
this alone will lead to a biased estimate of $\Pr(\win)$.
Second, our data preparation procedure may introduce another bias.
As discussed above,
we can only use teams where the identity of all five players is known,
but we cannot rule out that
these teams perform better in expectation than teams where
some players have kept their identities hidden.
For instance, perhaps players who have not opted-in to the data collection mechanism
are less experienced than those who have.
Several teams in our data
are from matches where the network of only that particular team was fully observed,
and the opposing team had some anonymous members.
This may cause $\Pr(\win)$ to further deviate from $0.5$.

To remove the effects of these (and possibly other unknown but similar) biases,
we use a bootstrap procedure \cite{efron1994introduction}
where we resample the data with replacement
so that $\Pr(\win) = 0.5$ in the resulting sample.
This is done simply by
sampling an equal amount of winning and losing teams,
and by making sure that
the total number of teams in resampled data
is equal to the size of the original data.
(As usual with bootstrap sampling,
this means that some teams can appear multiple times in the same sample.)
This is repeated 10000 times
to obtain 95\% bootstrap confidence intervals
for the estimates of the conditional probabilities $\Pr(\win\mid x = i)$ defined above.
We use the median of the bootstrap samples
as the final estimate of $\Pr(\win\mid x = i)$.

But considering these alone is not sufficient
to establish an association between
different conditions and team performance.
In particular,
as we partition the teams according to, for example $\mindeg$,
we obtain a slightly different probability estimate for every value of $\mindeg$,
even if there was no connection between $\mindeg$ and winning whatsoever.
The probability estimates will always to some extent
deviate from the baseline probability of $0.5$
due to statistical variation.
Some conditions specified by $\mindeg$ will always seem
to indicate a decrease or an increase in winning probability.
To draw robust conclusions
about the effects of network structure on performance,
the probabilities should be substantially higher or lower than $0.5$.
But when is a deviation large enough to be statistically significant?

We establish this by simulating the situation where
there explicitly is no connection between network structure and team performance.
That is,
we conduct another type of resampling on our data,
where the association between the independent variable
($\dens$, $\mindeg$, or $\maxdeg$)
and match outcome $\win$
{\em is explicitly removed}.
We do this by
permuting the $\win$ variable
within a dataset
uniformly at random across all teams
before estimating the winning probabilities.
This simple permutation procedure
is again repeated 10000 times to obtain
95\% {\em baseline confidence intervals} for
the estimates of $\Pr(\win\mid x = i)$
under the baseline
{\em where there is no association between
the winning probability and network structure}.

The justification for this is that
{\em if there is no association} between
some network statistic and winning the match,
then all permutations of $\win$ are {\em exchangeable}.
That is,
in a sense
the 10000 randomly chosen permutations of $\win$,
as well as the original observed one,
are equally likely to be ``true''.
This implies that the winning probabilities
estimated from the observed values of $\win$
should be similar to the ones estimated from
the 10000 permuted values of $\win$.
If, however, this is not the case,
meaning that the estimates from observed values are different
than the estimates from permuted values,
(e.g. consistently larger or smaller),
we have support for a hypothesis in which
there is an association between the networks statistic and winning.

Indeed, if the bootstrap estimate of $\Pr(\win \mid x = i)$
from the original non-permuted data
is outside of the basline confidence interval defined above,
we have support for the
hypothesis that network structure as specified by the condition $x=i$
has an effect on team performance.
The resampling procedure,
as well as all other analysis,
was implemented in R \cite{rproj}.
See \ref{app:resampling} for further technical details.


\section{Results}
\subsection{Descriptive statistics}
\textit{Skill level:} The teams are divided into three tiers in terms of player skill, called ``Normal Skill'' (1069 teams), ``High Skill'' (1151 teams), and ``Very High Skill'' (1948 teams). (There were 166 matches for which the skill specification is unavailable due to missing data. We made the simplifying assumption that teams in these matches are of ``Normal Skill''.)

\textit{Number of players:} The numbers of unique players in \radi,
\dire\ and \both\ are 10767, 9725, and 20192, respectively.
Table~\ref{table:player_freqs} shows the numbers of
players in each dataset broken down in terms of the number of matches they appear in.
We find that
only a handful of players appear in more than two matches.
This means that any variation we observe in performance is not caused by some particular players who differ from the others, e.g. have exceptional skill or behave maliciously.
\iffigure
\begin{table}
\centering
\caption{Number of players broken down in terms of the number of matches in which they appear.}
\label{table:player_freqs}
{\small
\begin{tabular}{rrrr}
number of matches & RADI & DIRE & BOTH \\
\hline
1 & 10561 & 9591 & 19573 \\
2 &  199  & 133  & 591   \\
3 &    7  &  1   & 27    \\
4 &    0  &  0   &  1    \\
\hline
\end{tabular}
}

\end{table}
\fi

\textit{Match duration:} The median match duration is 32 minutes with the 5th and 95th percentile at 30 and 35 minutes, respectively.

\textit{Baseline chances of winning:}
In all of our over 93k matches we find that Radiant wins with probability 0.584, and Dire with 0.416. This is in accordance with other studies, in which Radiant has been reported to have a higher winning probability. In the \radi, \dire\ and \both\ datasets the probabilities of Radiant to win are 0.623, 0.556, and 0.59, respectively.
Notice that by definition in every match in \radi, the Radiant team members are all non-anonymous, and their winning probability is $0.623 > 0.584$ (exact binomial test $p=0.00017$, 95\% CI$=[0.603, 0.644]$), and in \dire\ all Dire team members are known and Radiant wins with probability $0.556 < 0.584$ (exact binomial test $p=0.012$, 95\% CI$=[0.534,0.578]$). This suggests that teams where all players can be identified have a slightly higher chance of winning than teams with anonymous players.
Note that in the analysis below performance is estimated for teams with five non-anonymous members only.

\begin{figure}
\centering
\iffigure
\includegraphics[width=0.8\columnwidth]{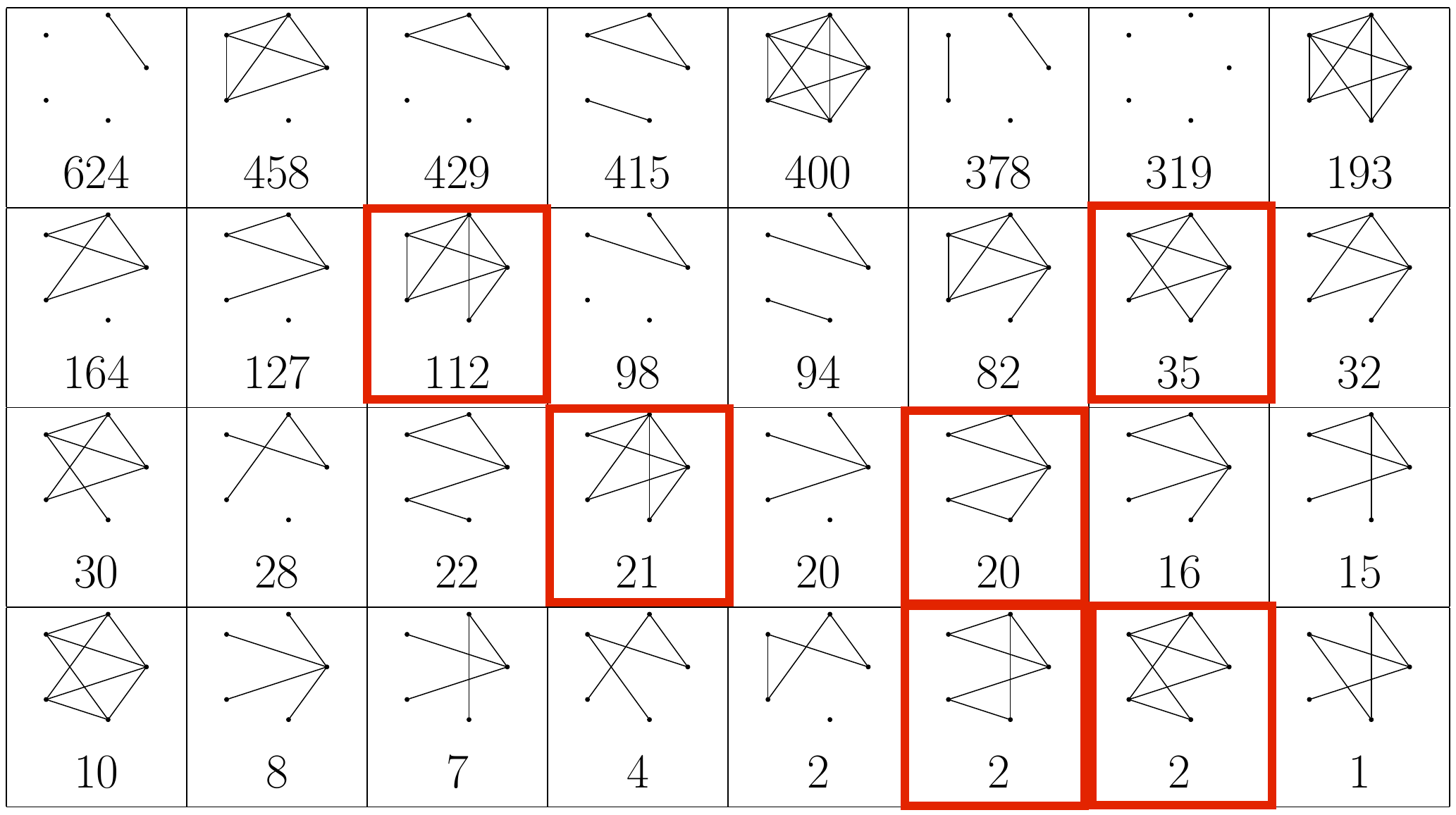}
\fi
\caption{Occurrence frequencies of different network configurations in dataset \both. Configurations with $\mindeg=2$ are marked with red boxes.}
\label{fig:isomorphisms}
\end{figure}

\textit{Network structure:} For teams of five players, the team
network has 34 different possible configurations\footnote{More formally, there are 34 isomorphism classes of undirected graphs that have five vertices each. Two graphs, $G_1$ and $G_2$, are isomorphic if we can rename the vertices of $G_1$ so that the resulting network is identical to $G_2$ without making any modifications to the edges.}, i.e., ways to assign connections between the players. Figure~\ref{fig:isomorphisms} shows 32 team configurations (out of the possible 34) as well as their occurrence frequencies observed in \both. Note that two of the possible configurations do not exist in the data at all.
In general the commonly occurring configurations tend to consist of a clique
plus some isolated nodes.
Also the completely disconnected team with five isolated nodes is one
of the more commonly occurring structures. Networks that are more
``random'' are also less frequent in our data. We want to point out
that network density is not correlated with occurrence frequency. The
frequent and infrequent configurations contain both dense and sparse
structures.
In the top-7 configurations (representing 3023 teams, or 72.5\% of all teams in \both) there are no networks with open triangles, i.e., sets of three nodes that contain two edges only, while all of the infrequent configurations have several open triangles. This is an expected outcome, as social networks such as the one used here tend to exhibit triadic closure \cite{granovetter1973strength},
meaning that ``two of my friends are most likely also friends with each other''.

\subsection{Network structure and performance}
\begin{figure}
\includegraphics[width=1.12\columnwidth]{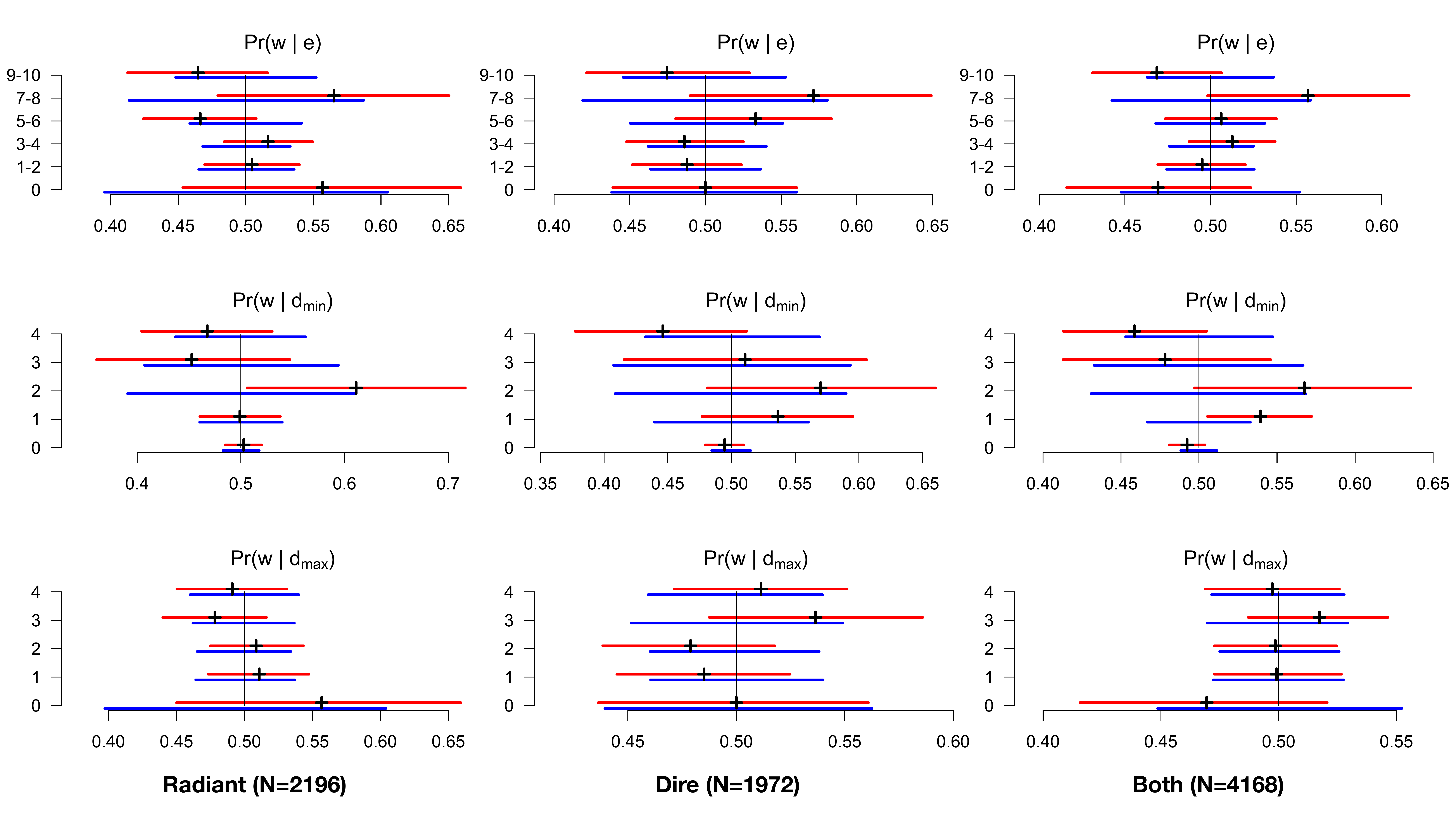}
\caption{Estimates of winning probability and the associated confidence intervals (red) in three datasets (columns) for three different network statistics (rows). The blue intervals show ranges of probable values of the estimate under a model where the association between network structure and winning has been explicitly removed.}
\label{fig:mainres}
\end{figure}
We proceed to describe our basic findings on
network structure and team performance.
The moderating role of team skill
is discussed later in Section~\ref{sect:skillstuff}.
Figure~\ref{fig:mainres} consists of nine panels,
each showing condition-specific estimates of
the winning probability (black plus signs)
together with a bootstrap confidence interval (in red),
as well as the ``null hypothesis'' confidence intervals (in blue)
that indicate the range of probable values of the estimate
when there is no association between network structure and performance.
The topmost row shows density,
the middle row shows minimum degree, and
the bottom row shows maximum degree.
Each column shows a different dataset.

We are especially interested in cases where the
probability estimate (black plus sign)
is outside (or very close to either end)
of the baseline CI (blue lines).
Moreover,
when comparing the estimates for two conditions
(of the same network statistic),
the bootstrap confidence intervals (in red)
should overlap as little as possible
if the claim is that
one of the conditions is associated with better performance than the other.
In general,
we refrain from using e.g.~$p$-values or other ``black-box'' statistics,
and rely on simple visual inspection of confidence intervals instead.
This is because we are mainly interested in finding
cases where any increases or decreases in winning probability
would translate to a concrete practical advantage.
That is,
small deviations are less interesting
even if they might be significant
in terms of our baseline model.

We observe a number of cases in Fig.~\ref{fig:mainres}
where the probability estimate of some condition
is low or high in relation to
what is expected given the baseline model:
\begin{enumerate}
\item $\Pr(\win \mid \dens = 5-6)$ is {\bf low} in \radi,             
\item $\Pr(\win \mid \mindeg = 2)$ is {\bf very high} in \radi,       
\item $\Pr(\win \mid \dens = 7-8)$ is {\bf high} in \dire,            
\item $\Pr(\win \mid \dens = 9-10)$ is {\bf low} in \both,            
\item $\Pr(\win \mid \dens = 7-8)$ is {\bf very high} in \both,       
\item $\Pr(\win \mid \mindeg = 1)$ is {\bf outside baseline CI} in \both,    
\item $\Pr(\win \mid \mindeg = 2)$ is {\bf very high} in \both, and    
\item $\Pr(\win \mid \mindeg = 4)$ is {\bf low} in \both.              
\end{enumerate}
As we have no reason to assume that
network structure and winning probability
would have different associations in \radi\ and \dire,
we can use them as controls for each other.
(I.e., this is a simple way of simulating a replication of the same experiment within the same study.)
Beginning with item 1 above
that suggests the condition $\dens = 5-6$
to have a low winning probability in \radi,
we find no similar effect in \dire,
where $\Pr(\win \mid \dens = 5-6)$ is in fact rather elevated.
However, for items 2 and 3
we find that the other dataset indicates a qualitatively similar effect,
albeit less strong.
In the combined dataset \both\
we observe five interesting cases (items 4--8),
two of which (items 5 and 7)
were also observed in either of the subsets and
corroborated by the other.
Observations in items 4, 6, 8 above, however, are
not suggested by either \radi\ or \dire\ alone,
and only become visible in the combined data.
Here we note that the magnitude of their effects,
i.e., the absolute decrease or increase in winning probability
is rather small (less than $0.05$ in all three cases),
even if residing outside the baseline CI as is the case for item 6.
Notably,
the maximum degree indicates no significant effects whatsover.
Also, \radi\ and \dire\ behave inconsistently for $\maxdeg$
suggesting that there indeed is no connection
between this statistic and team performance.

Taken together,
our main findings from Fig.~\ref{fig:mainres} are the following:
\begin{enumerate}
\item When considering {\bf density},
we find that there is a substantial difference
(as indicated by the bootstrap CIs shown in red)
in winning probability
between $\dens = 7-8$ and $\dens = 9-10$,
as well as between $\dens = 0$ and $\dens = 7-8$.
Performance is thus decreased
when the team network is
very sparse or a complete clique,
and best performance is found for a moderately connected network.
\item
When considering {\bf minimum degree},
we find that low values ($\mindeg = 1$ and $\mindeg = 2$)
are associated with substantially better performance than
higher minimum degrees ($\mindeg \geq 3$).
However, networks with isolated nodes ($\mindeg = 0$)
do not show increased performance.
This suggests that
not only should the team network as a whole
be ``moderately connected''
(as suggested by results on density),
but {\em each individual should also be
``moderately connected'' within the team}.
\item
We find no statistical support for
{\bf maximum degree} being associated with performance.
\end{enumerate}

\subsection{Team skill as a moderating variable}
\label{sect:skillstuff}
\begin{figure}
\includegraphics[width=1.12\columnwidth]{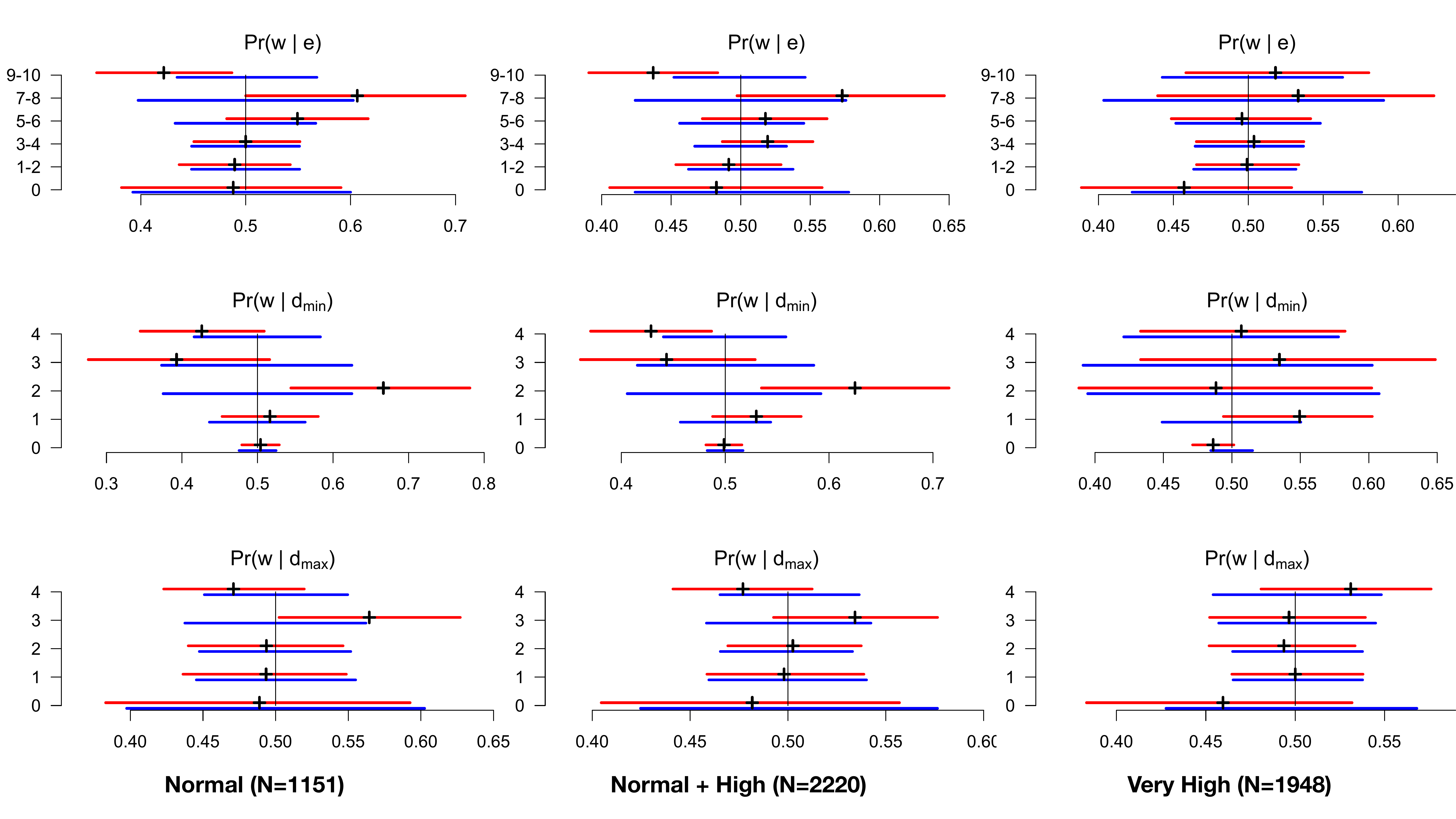}
\caption{Estimates of winning probability for different skill levels in dataset \both. We find that network structure is associated with performance for moderately skilled (``Normal'' and ``Normal + High'') teams, but for teams having ``Very High'' skill such effects are not present.}
\label{fig:skillres}
\end{figure}
It is conceivable
that in highly skilled teams,
factors such as previous familiarity with other members,
is not strongly associated with performance.
Our data seems to support this hypothesis.
Figure~\ref{fig:skillres} shows again nine panels
similar to the ones in Fig.~\ref{fig:mainres}.
This time
we only consider the dataset \both,
and have broken down the teams in terms of
the skill level they were assigned to by the game platform.
Fig.~\ref{fig:skillres} shows
``Normal'' skill teams in the leftmost column,
and ``Very High'' skill teams
in the rightmost column.
The middle column shows
a combination of ``Normal'' and ``High'' skill.

For teams with ``Normal'' skill,
we can observe that the estimate of winning probability
is outside the baseline CI
for the conditions
\begin{enumerate}
\item $\dens = 9-10$ (below baseline),
\item $\dens = 7-8$ (slightly above baseline),
\item $\mindeg = 2$ (substantially above baseline), and
\item $\maxdeg = 3$ (slightly above baseline).
\end{enumerate}
These findings are compatible with those
found above from Figure~\ref{fig:mainres} for all of \both,
but the effects are considerably stronger.
Most importantly,
we find that $\Pr(\win \mid \mindeg = 2)$ is clearly above $0.65$,
suggesting a substantial increase in
absolute magnitude of winning probability
(an increase over $0.15$ over the baseline of $0.5$)
for teams having moderately connected members.
Also the pairwise differences between $\mindeg = 2$
and the other conditions are significant
as indicated by the non-overlapping bootstrap CIs (shown in red).
Interestingly,
we observe a significant increase also for $\maxdeg = 3$
albeit the effect is rather small in terms of magnitude
(approx increase of $0.05$ over baseline of $0.5$).
This is further support for our finding that
in addition to the team being moderately connected as a whole,
{\em every member should also be moderately connected
at the individual level}.
Note that there are only  1151 ``Normal'' skill teams.
When ``Normal'' and ``High'' skill are considered together
(resulting in approximately twice as many teams as in ``Normal'',
middle column of Fig.~\ref{fig:skillres}),
we obtain essentially the same findings,
with minor differences in what conditions are
substantially outside the baseline CIs.

Finally,
when we only consider the ``Very High'' skill teams,
the picture is considerably different.
With a single exception, all estimates of the winning probability
for different conditions on density,
as well as minimum and maximum degree,
are clearly within the baseline confidence intervals.
The only difference is
the estimate for $\Pr(\win \mid \mindeg=1)$,
where we observe an almost significant and in absolute magnitude rather small increase
in comparison to the baseline.
However, unlike with ``Normal'' skilled teams
(including ``High'' skill),
we find {\em no support} for a robust association between
performance and network structure
for the ``Very High'' skill teams.

\section{Discussion}
Social network structure has been
highlighted as an important
factor of team performance.
Previous literature,
rather comprehensively summarised in \cite{balkundi2006},
has provided evidence for
network density being positively associated with team performance.
Some studies, such as work by \cite{OhEtAl2004} for example,
have also argued that ever increasing density
can be detrimental to performance.
In this work we
replicate these existing results
using novel, large-scale data from an online game.

Importantly,
we observe that
moderate density is associated with increased performance,
while complete cliques have substantially decreased performance.
As an important novel contribution
we consider team skill as a moderating variable,
and show that for teams having very high skill
there is essentially no association between network structure and performance,
something also suggested by \citet{balkundi2006},
while for teams of moderate skill such effects are present.

Another novelty of our approach
is the use of node degrees as a characterisation of network structure.
When team skill is not considered,
we do not in general find any strong associations between
the maximum degree of a network and team performance,
but we do find that networks where the minimum degree $\mindeg$ is equal to $2$
tend to exhibit substantially increased performance.
For teams of normal skill,
we in addition to the above result also observe a small but significant
increase for performance when the maximum degree $\maxdeg$ is equal to $3$.
While the precise values of $\mindeg=2$ and $\maxdeg=3$ are most likely specific
to our networks having exactly five members,
these results suggests that
{\em teams where every member is ``moderately connected''}
(in relation to the size of the team)
exhibit better performance.
Previous literature on team density
suggest that
as a whole,
a team should be moderately connected within itself,
and {\em our findings augment this with the observation
that the same holds for individual members of the team,
especially if the team is less skilled.}

We hypothesise that
our findings
can be partly explained by global properties of social networks.
As can be seen from Figure.~\ref{fig:isomorphisms}, networks with $\mindeg=2$ (marked by the red squares) tend to violate the {\em triadic closure property} \cite{granovetter1973strength}.
Such networks are more likely to have members from a number of different communities. In this case some members act as ``bridges'' by belonging to at least two of these communities, and are hence directly connected to members from both.
This type of network structure may introduce diversity into the teams that is beneficial for overall performance.

Numerous studies have argued that networks with the so called ``small-world'' property facilitate information diffusion within the network, and thereby increase the performance of the entire system irrespective of the underlying process \cite{watts1998collective,kleinberg2000navigation,latora2001efficient,cowan2004network,bassett2006small}. These studies, however, have focused on system-level phenomena in large networks, which may not directly carry over to smaller groups, such as teams.
Previous work on creative teams, in particular musical production teams and scientific collaborations, has suggested a connection between the structure of the global social network of the team members and success of the musical \cite{uzzi2005collaboration}, as well as the diversity of scientific teams in terms of incumbent members and newcomers and team performance \cite{guimera2005team}. These studies have also found that teams that have both familiar, as well as non-familiar members tend to perform better.
In this study we obtain similar results
by looking at the teams individually,
and only by considering the within team social networks.
Moreover, very recently similar results have also been reported for larger teams (14 members on average) in another multiplayer online game \cite{BenefieldSL16}.

\subsection{Limitations}
Our data is limited to small ad-hoc teams that are assembled on the fly to perform a relatively short, well-defined task.
However, we believe this is an important context for studying this problem, as for example knowledge work is increasingly carried out by independent freelancers in transient, information systems mediated teams.

The team networks were built with ``friend'' lists from the Steam social network. These are analogous to friendships in e.g.~Facebook, meaning that we can assume two connected team members to know each other at least to some degree, but the depth or longevity of their relationship remains unknown. 

Also, we observe and construct the team networks only after a match has been played. This is because we can retrieve the contact lists only once information about the match becomes public. Due to technical reasons this happens only a couple of hours after the match has taken place. Our data does not reveal if two players were disconnected during the match, but became connected in a period of a few hours after the match, or were connected during the match, but became disconnected later. However, there are no obvious reasons to assume that this would introduce a systematic bias to our observations, even if it may occasionally happen in our data.

Finally, the observation about performance decreasing for very high density networks may be an artefact of the way Dota 2, or similar online games in general are played.
For densely connected teams
the real source of enjoyment for players may simply be time spent with friends, and winning the match is secondary.
However,
it can also be argued that the converse is true:
a group of friends might play together often,
and hence employ finely tuned strategies and communicate more efficiently
that teams composed of strangers.

\subsection{Conclusion and future work}

While network density (or, equivalently, average degree) is a natural way to operationalise the overall cohesion or familiarity within a team,
the degree of a node captures a different phenomenon.
It focuses on an individual, and expresses the within-team connectivity from the perspective of a single team member.
Our results suggest that
taking this point of view may provide additional insights
to why certain network structures seem beneficial for team performance.
For example,
future studies may ask
if individuals who encounter both familiar as well as unfamiliar people in their teams
experience a stronger need to perform well to give a good impression
of their skills to the others?
And is this effect reduced when there are only familiar members?

\appendix
\section{Appendix}
\subsection{Additional details about Dota 2 matchmaking}
\label{app:dota2matchmaking}
For every match, players can either search for a match by themselves or with 1--4 Steam friends as a team (up to total of 5 players). The ``normal matchmaking'' algorithm first finds players or other partially filled teams with equal skill level to make a full team of 5 players. Then the matchmaking algorithm finds another team of equal total skill level and pits these two teams against each other. The teams move into a shared lobby, where they start picking their characters before the actual match begins. The game’s matchmaking algorithm pairs teams who have equal skill level with the expected winning likelihood of 50\% based on players so called matchmaking rating. Therefore, the skill level of individual players is to a large degree controlled for. As there are thousands of simultaneous Dota 2 players, the likelihood for not finding equally skilled players and teams is very unlikely.

\subsection{Additional details about Dota 2 game modes}
\label{app:dota2modes}
We focus on games played with the default game mode ``All
pick''. This sets the least amount of settings on the game, as well as teams' character
selection. Specialised rules could introduce opaque complexities into
how the game is played. In the ``All pick''-mode there are no restrictions
as to what characters the players can choose.

\subsection{Additional details about data scraping}
\label{app:scraping}
Dotabuff.com shows identity information only for players who have opted in to share their data. The identity of all 10 players who participated in a match is known only in a fraction of all observed matches, however. Some players are anonymous because they have chosen to opt-out of the data collection process. Of every non-anonymous player, the script we use for data scraping finds the identifier of their Steam profile on the page downloaded from dotabuff.com, and then proceeds to download their list of contacts from steamcommunity.com using this identifier. The Steam identifier is a unique, essentially random number that is assigned to every user account in the Steam system. We use the Steam profile identifier as the primary key to identify each player.

\subsection{Implementation details of the resampling procedure}
\label{app:resampling}
The combined bootstrap/permutation method
that we use is implemented as shown below:
\begin{enumerate}
\item Let $D$ denote the data we are resampling, let $n$ denote the size of $D$.
\item For $N = 10000$ times, do:
\begin{enumerate}
  \item Draw $n/2$ rows with replacement from those teams in $D$ that won ($\win = 1$).
  \item Draw $n/2$ rows with replacement from those teams in $D$ that lost ($\win = 0$).
  \item Combine the two samples drawn above to form a new dataset $D'$.
  \item Use $D'$ to compute $\Pr(\win\mid\dens = i)$ and $\Pr(\win\mid\mindeg = j)$ for all $i$ and $j$. These estimates are used to construct the bootstrap distribution of the observed winning probabilities. (I.e., the red lines in figures \ref{fig:mainres} and \ref{fig:skillres}.)
  \item Create a new dataset $\tilde D'$ that is a copy of $D'$, but where the $\win$ variable has been permuted uniformly at random across all observations. That is, we explicitly make sure that the connection between either $\dens$ or $\mindeg$ and $\win$ is broken.
  \item Use $\tilde D'$ to compute $\Pr(\win\mid\dens = i)$ and $\Pr(\win\mid\mindeg = j)$ for all $i$ and $j$. These estimates are used to construct the baseline confidence intervals of the winning probabilities. (I.e., the blue lines in figures \ref{fig:mainres} and \ref{fig:skillres}.)
\end{enumerate}
\end{enumerate}
This procedure results thus in two distributions: one over the non-permuted bootstrap samples, and another over the bootstrap samples where the $\win$ variable was subsequently permuted.

\bibliographystyle{apalike}
\bibliography{common/dotateams}




\end{document}